\documentclass[russian,reqno,dvips,12pt]{article} 
\usepackage[T2A]{fontenc}  
\usepackage[cp1251]{inputenc}  
\usepackage[russian]{babel}
\usepackage{epsfig,amsmath,amsfonts,verbatim,amssymb,amsthm}
\usepackage{longtable,lscape}

cm \textwidth=16.0 true cm \emergencystretch=10pt

\theoremstyle{plain}

\theoremstyle{remark}
\newtheorem{remark}{Remark}

\def\in{I}

\begin{document}

\title{\textbf{Table of stable chemical elements based on the ``intensity--compressibility factor'' diagram and on mean square fluctuations of energy and time }}

\author{\textbf{V.P. Maslov}
\thanks{ National Research University Higher School of Economics, Moscow, 123458, Russia;
Moscow State  University,   Physics Faculty,  Moscow, 119234, Russia
 }
}
\date{ }

\maketitle

\begin{abstract}

In this paper, a new  physical notion, intensity, is introduced. The notion of intensity occurs in a special statistics, known as Gentile statistics, which is asymptotically close to ordinary thermodynamics. The introduction of the new notion of intensity in the theory of nuclear matter essentially changes the thermodynamical picture. Moreover, we can say that the thermodynamics of nuclear matter is the antipode of standard thermodynamics.

On the basis of  the ``intensity--compressibility factor'' diagram and mean square fluctuations of energy and time, a new table of properties of stable chemical elements is obtained and presented in this paper.

\end{abstract}

\subsection*{The uncertainty relation in general form}

The uncertainty relation for the coordinate $p$ and the momentum $x$ in the form $\delta x \delta p \sim \hbar$,  where $\delta p$, $\delta x$ are the mean square deviations of the momentum operator $\hat p$ and the coordinate operator $\hat x$,  while $\hbar$  is the Planck constant, was obtained by Heisenberg in\cite{1} in  1927.

At the present time, the Heisenberg uncertainty relation is generally written in the form of the following inequality for the operators $p$ and $q$
\begin{equation}
	\delta q \delta p \ge \hbar/2.
\end{equation}
This inequality was obtained in 1927 by Kennard \cite{2}.

Three years later, Schr\"odinger \cite{3} and Robertson \cite{4} generalized inequality (1) to the case of an arbitrary pair of quantum observables $X$ and $Y$:
\begin{equation}\label{dxdy}
	\delta X \delta Y \ge\frac{1}{2} \sqrt{\left(\overline{XY-YX}\right)^2+\left(\overline{XY+YX}\right)^2}.
\end{equation}

The right-hand side of the last inequality contains the commutator, as well as the anticommutator, of the  operators $X$ and $Y$. Here the operators are assumed self-adjoint. These formulas reflect the Heisenberg uncertainty principle for the quantities appearing in the left-hand sign and determine their fluctuations.

As a rule, the Schr\"odinger equation is written in one of two forms: as an equation containing the time
$t$ and the self-adjoint operator $\hat{H}$
\begin{equation}
i  \hbar\frac{\partial}{\partial t}\Psi =\hat{H}\Psi,
\label{Shr-1}
\end{equation}
or in the form of a stationary equation (not involving time) which sets the spectrum problem for the operator $\hat{H}$
\begin{equation}
\Lambda\Psi=\hat{H}\Psi.
\label{Shr-2}
\end{equation}

It is natural to relate the operator  $\hat{H}$ with the eigenvalue $\Lambda$, or , as physicists say, indicate the time and the phase.

Since the Cauchy problem is considered for the Schr\"odinger equation, i.e., the initial-value problem with
$t=0$, if follows that time changes from 0 to plus infinity and the operator   $i\hbar\frac{\partial}{\partial t}$   cannot be self adjoint.

In quantum physics, since the time of von Neumann and Pauli, the notion of observable quantity has been considered and put in correspondance with a self-adjoint operator in Hilbert space. Von Neumann and Stone considered only self-adjoint operators and observable quantities.

One of the difficulties in the standard formulation of quantum mechanics is in the impossibility of assigning to such quantities as time (Pauli), phase (Dirac), angle, and so on, an appropriate operator in the Hilbert space of the system. However, such difficulties can be avoided by taking for observable quantities nonorthogonal partitions of unity subjected to covariance constraints that generalize the Weyl commutation relations in the Stone--von Neumann theorem.

A.S. Holevo in \cite{5} and other papers was the first to propose a mathematical approach which allowed to assign  observable quantities to non-self-adjoint operators. The general outline of his approach is to consider   partitions of unity forming a convex set satisfying the covariance conditions and to distinguish the extreme points of this set that minimize the uncertainty functional in a certain state. In this way, one can obtain generalized observables, canonically corresponding to achievement time. In spectral theory, nonorthogonal partitions of unity arise, in particular, as generalized spectral measures for non-self-adjoint operators. In this context, for example, the operator corresponding to observable time turns out to be maximal Hermitian (but not self-adjoint). Using this approach, one can also show that relativistic massless particles turn out to be ``approximately localized'' if, in the definition of localization, arbitrary partitions of unity are allowed \cite{6},~\cite{7}.

From the contemporary point of view, the standard notion of observable corresponds to ``sharp'' observables, while the nonorthogonal partition of unity describes ``unsharp'' observables. In this situation, one can assign a probability distribution to sharp as well as unsharp observables in any state, which allows to calculate all the statistical characteristics -- mean values, dispersion, correlation, and to establish the uncertainty relation for nonstandard canonically conjugate pairs~\cite{7},~\cite{8}.

The time operator and the corresponding uncertainty relation has been studied by other authors, but the approach from the point of view of covariant observables appears to be most natural.

In the work of Olkhovsky~(see \cite{9} and other papers) based on Holevo's work, an approach convenient for physicists was developed. It allows to obtain the uncertainty relation for the time and energy operators in the form
\begin{equation}\label{dedt}
	\delta E \delta t  \geq \hbar/2.
\end{equation}

 In particular, Olkhovsky considered the Yukawa potential and  the BBGKY hierarchy (Bogoliubov--Born--Green--Kirkwood--Yvon hierarchy).
 In the paper~\cite{10}, an explicit analytic formula for the S-matrix in the case of an arbitrary central interaction inside a sphere of finite radius with the ``tail'' of the  Yukawa potential at large distances was obtained. This method uses the completeness of the space of wave functions outside a finite sphere, as well as the unitarity and symmetry of S-matrices.

In the present paper, we shall consider the uncertainty relation for time and energy. The problem of introducing the time operator in quantum mechanics goes back to its very origins. The time operator
$\hat t$, defined as
 \begin{equation}\label{top}
 \begin{array}{c}{\widehat{t}=t, \quad \text {time-like} t \text { -representation, }} \\
 {\widehat{t}=-i \hbar \frac{\partial}{\partial E}, \quad \text { energy-like $E$-representation, }}
\end{array}
\end{equation}
is not self-adjoint \cite{11}, but is Hermitian. It is precisely for this reason that V. Pauli, one of the founders of quantum mechanics, refused to use the time operator given by~\eqref{top}.

\subsection*{Passage from Bose to Fermi }

Consider the case in which at every energy level $E_p$, there can be no more than $K$ particles (so called Gentile statistics, or parastatistics). For an ideal gas of dimension $D=3$ the following relations for the number of particles $N$ and the energy $E$ hold:
\begin{equation}\label{Gent}
	N=\frac{V}{\lambda^{3}}(\operatorname{Li}_{3/2}(\in) -\frac{1}{(K+1)^{1/2}}\operatorname{Li}_{3/2}(\in^{K+1})) ,
\end{equation}
\begin{equation}\label{Mp}
	{E}= \frac{3}{2} \frac{{V}}{\lambda^{3}}T(\operatorname{Li}_{5/2}(\in)-\frac{1}{(K+1)^{3/2}}\operatorname{Li}_{5/2}(\in^{K+1})),
\end{equation}
where $V$ is the volume,  $T$ is the temperature,  $\operatorname{Li}_{(\cdot)}(\cdot)$ is the polylogarithmic function,
$\in$ is a new  quantity that we will call
{\it intensity} and $\lambda$ is the de Broglie wavelength:
\begin{equation}\label{lam}
	\lambda=\sqrt{\frac{2\pi\hbar^2}{m T}};
\end{equation}
here $m$ is the mass.

Gentile statistics, which coincides with the notion of polylogarithm and based on the formula
$\operatorname{Li}_{s}(z) =\sum_{k=1}^\infty\frac{z^k}{k^s}$, is related to the Bose--Einstein and Fermi--Dirac
distributions in the following way.
The integral of the  Bose-Einstein distribution is  expressed in terms of the polylogarithm as:
$$
\operatorname{Li}_{s+1}(z) =\frac{1}{\Gamma(s+1)} \int_0^\infty \frac{t^s}{e^t/z-1}\, dt.
$$
The integral of the  Fermi-Dirac distribution is  also expressed in terms of the polylogarithm:
$$
\operatorname{Li}_{s+1}(-z) =\frac{1}{\Gamma(s+1)} \int_0^\infty \frac{t^s}{e^t/z+1}\, dt.
$$
Therefore, the passage  from  the  Bose-Einstein distribution  to the  Fermi-Dirac distribution
corresponds to the passage,
in the Hougen--Watson $P-F$ diagram  ($P$  is  the pressure, $F=PV/(NT)$   is the  compressibility  factor),
from the quadrant  II ($P<0, F>0$) to the quadrant  IV  ($P>0, F<0$).

Values $s$   in relations for   polylogarithm  can be arbitrary. However in the case  when   only fixed  values (e.g. $s=2$, $s=3$, etc) are given, one can consider  all other  values  of $s$   as  holes.  Holes  are the  values which we leave out.

 The polylogarithm function sometimes appears in thermodynamics.
However, the polylogarithm, as well as Gentile statistics, have no direct relationship to classical thermodynamics,
which includes temperature, chemical potential, energy, lifetime, entropy, and other parameters used in classical
and quantum thermodynamics. From the purely mathematical viewpoint, as well  as asymptotically,
the relations of Gentile statistics and of thermodynamics are very similar.
 But it should be stressed that Gentile statistics and the notion of polylogarithm
 have no relationship to thermodynamics.

Nevertheless, we introduce, as a new physical thermodynamical quantity, a certain constant related to Gentile statistics and to the
polylogarithm. We have called this constant the intensity and denote it by $\in$.  It is close to the notion of activity $a=e^{\mu/T}$
in standard thermodynamics, but certainly does not coincide with it. Basically, the new parameter $\in$ is related to the notion of   ``lacunary indeterminacy''~\cite{12}, in situations when not all quantities are exactly defined.  In~\cite{12}, the meaning of lacunary indeterminacy was  explained as follows:   in a certain region we cannot distinguish and number particles,  however we can determine the  total number of particles. This region is a lacuna of sorts in the general deterministic picture.

In order to relate the  Bose-Einstein and Fermi-Dirac distributions with  the principle of lacunary indeterminacy,
it  is necessary to construct an analog of Mendeleev's periodic table of elements for stable gases
on the basis of experimental data. In the present paper, we present such a table for  stable chemical elements;
this table is related to lacunary indeterminacy and, except for the parameter
$\in$, has nothing to do with Gentile statistics.

The table presented by the author is indeed related to lacunary indeterminacy, but does not pretend to explain any law of nature, just as Mendeleev's table, which initially was purely empirical and gave no explanation of all the  physical phenomena that lead to it (see, for example, \cite{17}).

The passage of particles from the domain where the Bose--Einstein statistics is obeyed to that where the Fermi--Dirac statistics rules, goes through a domain known as a fur coat or a  region of  lacunary indeterminacy~\cite{12}. The minimal value of the intensity  $\in$,  for which the number of Bose particles $N$ tends to $0$, will be denoted by $\in_0$. The quantity $\in_0$ shows for what minimal intensity the decomposition of a boson into fermions begins.

In our previous papers~\cite{12}--\cite{14},  we have obtained an expression for   $\in_0$, i.e., for those values of~ $\in$ for which $K=N=0$:
\begin{equation}
	\label{N=0}
	\frac{1}{2} \operatorname{Li}_{3/2}(\in_0)-  \log (\in_0)\operatorname{Li}_{1/2}(\in_0)-B^{-1}=0,
\end{equation}
where $B=\frac{V}{\lambda^{3}}>0$.

 \begin{figure}
	\centering
	\includegraphics[width=14cm]{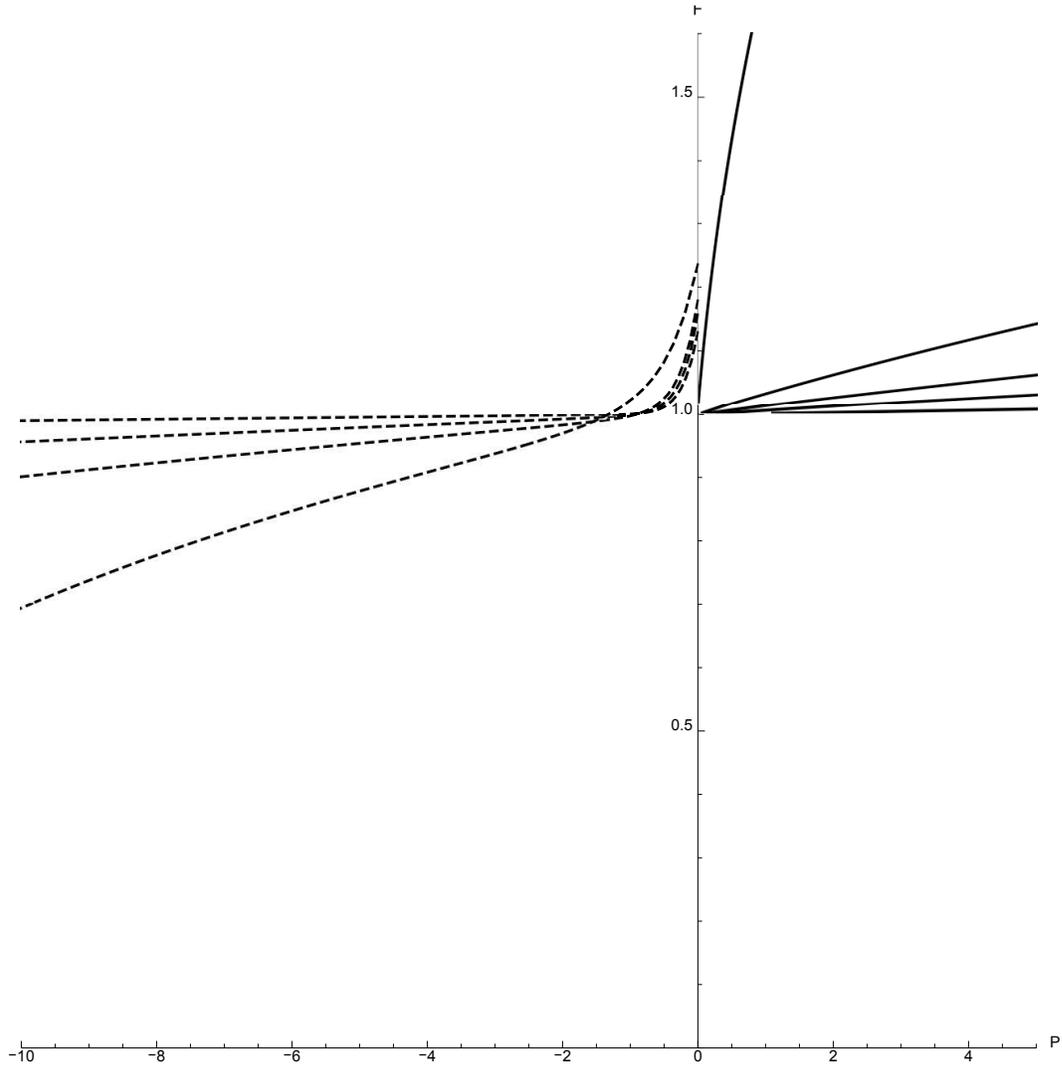}
	\caption{Dependence of the compressibility factor $F=PV/(NT)$ on the pressure $P$ expressed in $MeV/fm^3$ for helium-4, litium-6, litium-7, berillium-9 (from top to bottom along the $F$ (vertical) axis). The continuous lines represent the Fermi branch. The hashed lines are isotherms of the Bose branch, constructed according to formulas~\eqref{Gent} - \eqref{Mp}.}\label{key}
	\label{fig_01}
\end{figure}

\begin{figure}
	\centering
	\includegraphics[width=8.6cm]{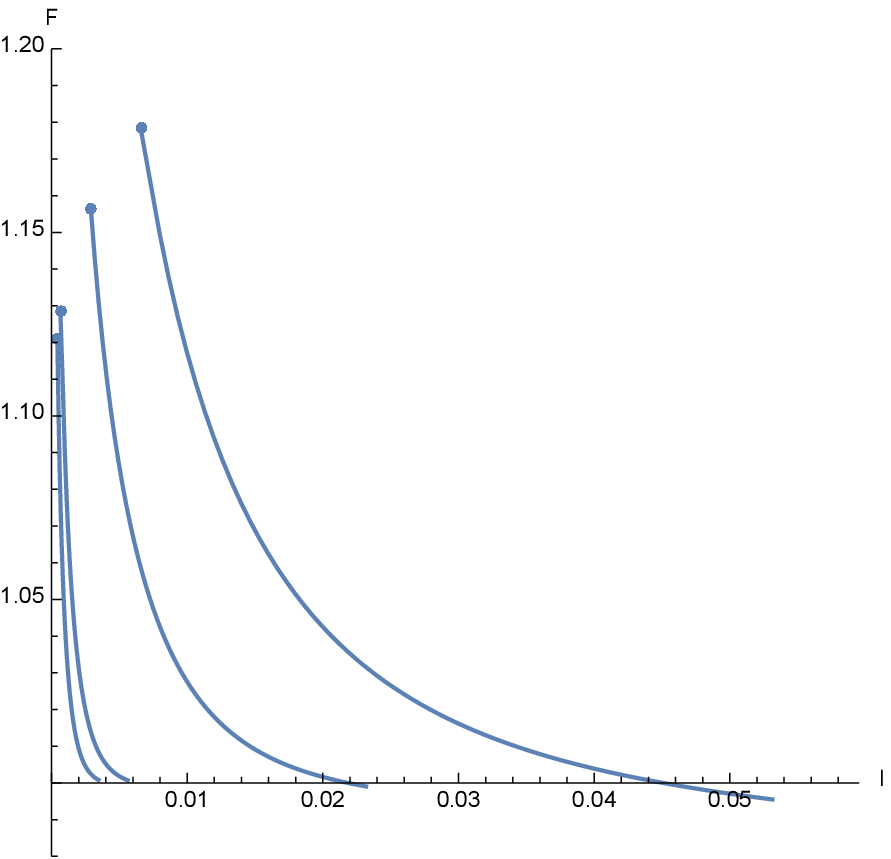}
	\caption{ Dependence of the compressibility factor $F$ on the intensity  $\in$, for berillium-9, litium-7, litium-6, helium-4 (from left to right). The curves are the isotherms of the Bose branch and are plotted according to formulas~\eqref{Gent} -- \eqref{Mp}. The temperature is equal to the  binding energy $E_b$ of the nucleus (see Table~\ref{tabl:t1}).}
	\label{fig_02}
\end{figure}

In thermodynamics, it is customary to use the Hougen--Watson P-F diagram ($P$ is the pressure, $F=PV/(NT)$ is the compressibility factor), which illustrates the Van der Waals law of corresponding states. On the P-F diagram,  the Fermi--Dirac distribution corresponds to the domain of positive values.
 Figure~\ref{fig_01} represents the P-F diagram for helium-4, litium-6, litium-7, and berillium-9.   .

At this point, the author's approach consists in using an analogous diagram, the $\in$-F diagram ($\in$ is the intensity, $F$, the compressibility factor), for very high temperature isotherms. This diagram is in a certain sense the antipode of the P-F diagram. Figure~\ref{fig_02} shows the Bose branch on the $\in$-F diagram for the same chemical elements.

In the paper \cite{14},  it was established that  in the passage from the Bose branch to the Fermi branch a jump in the specific energy $E_{sp}$ occurs, and the value of this jump is given by the formula
\begin{equation}\label{DE}
\Delta E_{sp}=T(\gamma+1)(\left. F \right|_{\in=\in_0}-1),
\end{equation}
where $\gamma=D/2-1$.

According to the formula
\begin{equation}\label{Dt}
\delta t_{min}=\hbar/(2\Delta E_{sp})\ge \delta t,
\end{equation}
we can compute the minimal interval of time $\delta t_{min}$ required to discover the energy jump $\Delta E_{sp}$. If we introduce the time operator, we can consider its mean square fluctuation $\delta t$. Then $\delta t_{min}$ is the minimal time fluctuation corresponding to the energy jump $\Delta E_{sp}$.

Let us use the approach due to Kvasnikov~\cite{15}  and show how one can find the mean square fluctuation for the number of particles $N$ in the case when $N$ is infinitely small.

\subsection*{Passing to infinitely small quantities}

From now on, we shall denote the mean square value of the number of particles by $N$, omitting the bar. Let us find the value of
$ \left(\frac{\partial \overline{N}}{\partial \mu}\right)_{TV}$,
which is needed  to compute $N$. We shall assume that $K=N$. Let us introduce the notation
\begin{equation}
	\phi(\mu,N)=\left(\frac{{2 \pi mT }}{(2 \pi \hslash)^2}\right)^{\gamma+1}V(\text{Li}_{1+\gamma}(\in)-\frac{1}{(N+1)^{\gamma}}\text{Li}_{1+\gamma}(\in^{N+1}))- N.
\end{equation}

As was shown in~\cite{12}, there is a one-to-one correspondence between the chemical potential $\mu$ and the number of particles $N$. Let us expand the function  $\phi(\mu(N),N)$ into a Maclaurin series in the variable $N$
\begin{equation}\label{phn}
	\begin{split}
		&\phi(\mu_0,0)+N[ \phi_{\mu}(\mu_0,0)\mu_N+\phi_{N}(\mu_0,0)]\\
		&+N^2[ \phi_{\mu\mu}(\mu_0,0)\mu_N^2+\phi_{\mu}(\mu_0,0)\mu_{NN}+2\phi_{\mu N}(\mu_0,0)\mu_N+\phi_{N N}(\mu_0,0)]+O(N^3),
	\end{split}
\end{equation}
where the derivative of  $\mu_N$ is computed at the point  $\mu=\mu_0$.

 Since $\phi(\mu,0)\equiv0$, the quantities $\phi(\mu_0,0),$
 $\phi_{\mu}(\mu_0,0)$ and $\phi_{\mu\mu}(\mu_0,0)$ are all equal to zero. Dividing by $N$, we obtain the equality
 \begin{equation}
 	\label{phn1}
\phi_{N}(\mu_0,0)+N[2\phi_{\mu N}(\mu_0,0)\mu_N+\phi_{N N}(\mu_0,0)]+O(N^2)=0.
 \end{equation}
Let us compute $\phi_{N}(\mu_0,0)$:
 \begin{equation}\label{mu0}
	\phi_{N}(\mu_0,0)=\left(\frac{{2 \pi mT }}{(2 \pi \hslash)^2}\right)^{\gamma+1}V[ \gamma  \text{Li}_{\gamma+1}(\exp[\mu_0/T])-  \log (\exp[\mu_0/T]) \text{Li}_\gamma(\exp[\mu_0/T])]-1
\end{equation}

The value of $\mu_0$ is chosen so as to have  $\phi_{N}(\mu_0,0)=0$ (see formula \eqref{N=0} for the calculation of $\in_0$). If we now pass to the limit as  $N\to0$ in the expression \eqref{phn1}, then the derivative
 \begin{equation}\label{mun0}
\mu_N=-\frac{\phi_{N}(\mu_0,0)}{ \phi_{\mu}(\mu_0,0)}.
\end{equation}
will have an uncertainty of type <<0/0>> and, therefore, cannot be calculated according to this formula.

Dividing by $N$ once again, we obtain
 \begin{equation}\label{2ph}
2\phi_{\mu N}(\mu_0,0)\mu_N+\phi_{N N}(\mu_0,0)+O(N)=0.
\end{equation}
The passage to the limit as  $N\to0$ in \eqref{2ph}  gives the value of the derivative  $\mu_N$ at the point $\mu=\mu_0$
 \begin{equation}\label{mun}
\mu_N=-\frac{\phi_{N N}(\mu_0,0)}{2\phi_{\mu N}(\mu_0,0)}.
\end{equation}

The expressions for the partial derivatives
 $\phi_{N N}(\mu_0,0), \phi_{\mu N}(\mu_0,0)$ are of the form
 \begin{equation}
\phi_{N N}(\mu_0,0)=-\left(\frac{{2 \pi mT }}{(2 \pi \hslash)^2}\right)^{\gamma+1}V[\log ^2(\in_0) \text{Li}_{\gamma-1}(\in_0)+\gamma ((\gamma+1) \text{Li}_{\gamma+1}(\in_0)-2 \log (\in_0) \text{Li}_\gamma(\in_0))]
\end{equation}

 \begin{equation}
	\phi_{\mu N}(\mu_0,0)=-\frac{1}{T}\left(\frac{{2 \pi mT }}{(2 \pi \hslash)^2}\right)^{\gamma+1}V[(1-\gamma)  \text{Li}_{\gamma }(\in_0)+\log (\in_0) \text{Li}_{\gamma -1}(\in_0)]
\end{equation}

Substituting the last two expressions into formula  \eqref{mun}, leads to a formula for the derivative at the point $\in=\in_0$
\begin{equation}\label{dnmuas0}
	\left(\frac{\partial \mu}{\partial N}\right)_{TV}
	=-\frac{T}{2}\frac{\log ^2(\in_0) \text{Li}_{\gamma-1}(\in_0)+\gamma ((\gamma+1) \text{Li}_{\gamma+1}(\in_0)-2 \log (\in_0) \text{Li}_\gamma(\in_0))}{ (1-\gamma)  \text{Li}_{\gamma }(\in_0)+\log (\in_0) \text{Li}_{\gamma -1}(\in_0)}
\end{equation}
and to an expression for the dispersion of the number of particles of the system
\begin{equation}\label{dnmuas}
	\overline{(\Delta N)^{2}}=-{2}\frac{ (1-\gamma)  \text{Li}_{\gamma }(\in_0)+\log (\in_0) \text{Li}_{\gamma -1}(\in_0)}{\log ^2(\in_0) \text{Li}_{\gamma-1}(\in_0)+\gamma ((\gamma+1) \text{Li}_{\gamma+1}(\in_0)-2 \log (\in_0) \text{Li}_\gamma(\in_0))}.
\end{equation}

Recall that for an arbitrary quantity $x$, its mean square fluctuation is defined as $\delta x=\sqrt{\overline{(\Delta x)^{2}}}.$
Using the well known uncertainty relation $\delta N \delta \mu\ge T$ for grand  canonical ensembles (see~\cite{16}), we can find the value of the minimal admissible mean square fluctuation
\begin{equation}\label{heismu}
 \delta \mu_{min}=T/{\delta N}\le \delta \mu.
\end{equation}

\subsection*{Table of stable nuclei}

We obtained Table~\ref{tabl:t1} of stable nuclei of chemical elements using the data base  IsotopeData, included in the software  Wolfram Mathematica and containing 255 stable elements.

In Table ~\ref{tabl:t1}, the values for the fluctuation of energy and time were obtained by means of formulas  ~\eqref{DE} and  \eqref{Dt}, respectively.  The time $t$ and the intensity $\in$ are related by the uncertainty equation.

Thus, the intensity $\in$ plays the main role in this table. If the value of    $\in_0$  is sufficiently small, the corresponding temperature (and the energy) will be huge.  If the quantity $\in_0$ is of the order of 1, then the temperature (and the exitation energy) will be small.

 \begin{remark}
In the table, we have not taken in consideration the tunnel effect, which can substantially affect the stability of nuclei.
 \end{remark}

In Table~\ref{tabl:t1}, the following parameters of stable isotopes of different nuclei are presented:
\begin{itemize}
	\item{\text{No} {\textnumero} -- the number of the nucleus in our list}
	\item{ isotope -- the name of the corresponding chemical element, with its mass number $A$ (i.e., the number of nucleons in the nucleus) after the hyphen}
	\item{$Z$ -- the charge number (i.e., the number of protons in the nucleus)}
	\item{$\in_0$ -- the minimal value of the intensity $\in$ for Bose particles, calculated according to formula  \eqref{N=0}}
      \item {$E_b$ -- the binding energy, which is equal to the temperature $ T $ of the nucleus}
      \item{$\Delta E_{sp}$ -- the jump in the specific energy in the passage from the Bose branch to the Fermi branch (formula \eqref{DE})}
      \item{$\delta t_{min}=\hbar/(2\Delta E_{sp})$ -- time fluctuation (in seconds)}
      \item{$\delta N$ -- mean square fluctuation of the number of particles equal to  $\sqrt{\overline{(\Delta N)^2}}$ and computed at the point  $\in_0$ according to formula  \eqref{dnmuas}}
      \item{$\delta \mu_{min}$ -- minimal mean square fluctuation of the chemical potential at the point $\in_0$, equal to  $T/{\delta N}$ }	

\item{$\mu_0$ -- chemical potential $\mu_0 = T \log {\in_0} $}

\item{$F(\in_0)$ -- value of the compressibility factor at the point  $\in=\in_0$}

\end{itemize}

The value of $\in_0$ in the case of the disintegration of the nucleus is found from formula \eqref{N=0} taking into account the expression for the de Broglie wavelength $\lambda$, calculated from the known volume $V$ of the nucleus, the temperature$T$ of the nucleus and the mass $m$. The volume of the nucleus is taken to be equal to the volume of a ball of radius
$r_0=A^{1/3} 1.2\times 10^{-15}$m. The temperature $T$ of the nucleus expressed in energy units, is taken to be equal to the binding energy $E_b$ of the nucleus (taken from the database IsotopeData). In the table all the quantities having the dimension of energy are given in MeV units. The mass $m$ is taken to be the mass of the whole nucleus. In all the calculations, the three-dimensional case was considered, i.e.,$\gamma=D/2-1=1/2$.

In conclusion, let us note that the notion of intensity $\in_0$ introduced by the author is a new quantity related to Gentile statistics,
to the notion of infinitely small quantity, and to the polylogarithm function. This quantity bears no relationship to thermodynamics and statistical physics, just as the $\in_0-F$ diagram has no relationship to ordinary thermodynamics. We have obtained this new
thermodynamics by transforming ordinary thermodynamics into its ``antipode''. An example of the antipode
of a physical quantity can be the same quantity taken with a minus sign. Thus, the antipode of a particle is an antiparticle.

The notion of antiparticle was introduced speculatively. There was nothing in reality corresponding to it until physicists gave it the meaning of a hole, i.e., the absence of a particle. Similarly, before the author's recent papers, the notion of ``infinitely small number of particles'' did not exist. This important notion is part of the antipode of standard thermodynamics. It allows to develop nuclear physics in connection with nuclear matter.

\newpage
\begin{landscape}
{\small
{

\begin{longtable}{|c|c|c|c|c|c|c|c|c|c|c|}
	\caption{Parameters of stable isotopes of different nuclei} \label{tabl:t1} \\
	\hline
	\hline
	\textnumero&isotope&$Z$&$\in_0$& $E_b$  &$\Delta E_{sp}$&$\delta t_{min}$&$\delta N$&$\delta {\mu}_{min}$ & $\mu_0$&$F(\in_0)$ \\
	\hline
	\endfirsthead
	\hline
	\textnumero&isotope&$Z$&$\in_0$& $E_b$  &$\Delta E_{sp}$&$\delta t_{min}$&$\delta N$&$\delta {\mu}_{min}$ & $\mu_0$&$F(\in_0)$ \\
	\hline
	\endhead
	\hline
	\endfoot
	\hline
	\endlastfoot
		1   & lead-208         & 82 & 1.636*10\textasciicircum{}(3) & 1.1341*10\textasciicircum{}(2) & 2.9018*10\textasciicircum{}(-24) & 2.9672*10\textasciicircum{}(-1) & 5.515*10\textasciicircum{}(3)  & 6.57*10\textasciicircum{}(-10) & -3.46*10\textasciicircum{}(4) & 1.046203 \\
		2   & lead-207         & 82 & 1.629*10\textasciicircum{}(3) & 1.13*10\textasciicircum{}(2)   & 2.9123*10\textasciicircum{}(-24) & 2.9685*10\textasciicircum{}(-1) & 5.488*10\textasciicircum{}(3)  & 6.7*10\textasciicircum{}(-10)  & -3.44*10\textasciicircum{}(4) & 1.046245 \\
		3   & lead-206         & 82 & 1.622*10\textasciicircum{}(3) & 1.1264*10\textasciicircum{}(2) & 2.9218*10\textasciicircum{}(-24) & 2.9697*10\textasciicircum{}(-1) & 5.463*10\textasciicircum{}(3)  & 6.83*10\textasciicircum{}(-10) & -3.42*10\textasciicircum{}(4) & 1.046286 \\
		4   & thallium-205     & 81 & 1.615*10\textasciicircum{}(3) & 1.1224*10\textasciicircum{}(2) & 2.9322*10\textasciicircum{}(-24) & 2.971*10\textasciicircum{}(-1)  & 5.436*10\textasciicircum{}(3)  & 6.96*10\textasciicircum{}(-10) & -3.41*10\textasciicircum{}(4) & 1.046328 \\
		5   & mercury-204      & 80 & 1.609*10\textasciicircum{}(3) & 1.1189*10\textasciicircum{}(2) & 2.9413*10\textasciicircum{}(-24) & 2.9723*10\textasciicircum{}(-1) & 5.412*10\textasciicircum{}(3)  & 7.1*10\textasciicircum{}(-10)  & -3.39*10\textasciicircum{}(4) & 1.046369 \\
		6   & lead-204         & 82 & 1.608*10\textasciicircum{}(3) & 1.1181*10\textasciicircum{}(2) & 2.9433*10\textasciicircum{}(-24) & 2.9723*10\textasciicircum{}(-1) & 5.408*10\textasciicircum{}(3)  & 7.11*10\textasciicircum{}(-10) & -3.39*10\textasciicircum{}(4) & 1.046372 \\
		7   & thallium-203     & 81 & 1.601*10\textasciicircum{}(3) & 1.1145*10\textasciicircum{}(2) & 2.9528*10\textasciicircum{}(-24) & 2.9736*10\textasciicircum{}(-1) & 5.384*10\textasciicircum{}(3)  & 7.25*10\textasciicircum{}(-10) & -3.37*10\textasciicircum{}(4) & 1.046414 \\
\end{longtable}
}
 }

\end{landscape}
I express my deep gratitude to E.I. Nikulin for help in the computations and in the plotting of figures.

The paper was written in the framework of the State Assignment (No. AAAA-A17-117021310377-1).

\section*{Additional information}

\textbf{Competing  interests:}  The author declares no competing  interests.

\end{document}